 \documentstyle[12pt]{article}
 \newcommand{\eps}{\epsilon}

\begin{document}
 \renewcommand{\theequation}{\thesection.\arabic{equation}}

\begin{titlepage}

\begin{center}

{\Large \bf Three questions on Lorentz violation\footnote{Invited
lecture at DICE2006 Piombino - Italy, September 11-15, 2006}}

\vspace{2.5cm}

{\large Alfredo Iorio}$^{a, b}$

\vspace{.5cm}

{\sl $^a$ Institute of Particle and Nuclear Physics, Charles University} \\
{\sl  V Holesovickach 2, 180 00 Prague 8 - Czech Republic}

{\sl $^b$ Department of Physics, University of Salerno} \\
{\sl  Via Allende, 84081 Baronissi (SA) - Italy}

\vspace{.5cm}

%\noindent E-mail {\rm iorio@ipnp.troja.mff.cuni.cz}

\vspace{2.5cm}

\begin{abstract}
We review the basics of the two most widely used approaches to
Lorentz violation - the Standard Model Extension and
Noncommutative Field Theory - and discuss in some detail the
example of the modified spectrum of the synchrotron radiation.
Motivated by touching upon such a fundamental issue as Lorentz
symmetry, we ask three questions: What is behind the search for
Lorentz violation? Is String Theory a physical theory? Is there an
alternative to Supersymmetry?
\end{abstract}
\end{center}

\end{titlepage}

\section{Introduction}

The Special Theory of Relativity (STR) is again the subject of an
intense investigation. The first time such a scrutiny took place
was about one hundred years ago, when the theory was first
proposed by Einstein. At that time STR represented a brilliant
theoretical solution to many crucial experimental questions: the
experiments - for instance those on the ``luminiferous aether'' -
were calling for a theoretical revolution, that more or less
punctually came.

It is interesting to read how eminent scientists of the time
reacted to STR. For instance, it is quite noticeable the heroic
effort of Bridgman - an experimentalist and Nobel laureate -  to
avoid to future physicists (us) another cultural shock of the kind
that struck scientists with STR - and Quantum Mechanics (QM) - in
the early 1900's \cite{bridgman}. Such an effort, although not
really successful regarding the main goal of Bridgman, gave origin
to a method of philosophical investigation on the meaningfulness
of scientific theories that today goes under the name of {\it
operationism}. It is surprising to learn that even Mach, who with
his critics to mechanicism paved the way on the philosophical side
to STR, was skeptic with Einstein's theory \cite{mach}. But
skeptics had to surrender to the enormous amount of experimental
evidences from a great variety of different kind of phenomena.
Thus STR is probably the most solidly tested and deeply rooted
theory in contemporary particle physics.

Despite STR's experimental robustness, nowadays is happening
somehow the reverse of what happened at the beginning of the last
century: theoretically posited questions eagerly call for
experimental findings that {\it violate} STR. That a theory calls
for experiments to prove (or disprove) its assertions it is
certainly not new in physics. To a great extent already STR, and
more evidently the General Theory of Relativity (GTR), are
examples of this. We may even say that it is the goal of any good
theoretical physicist to produce a theory that predicts unknown
phenomena while keeping explaining the known ones.

Nonetheless, the search going on these days for experimental
signatures of violation of Lorentz symmetry motivates, in our
view, some general reflections on contemporary ``theoretical
particle physics'', i.e. of a large - perhaps the dominant - part
of what on the Los Alamos archives goes under the name of HEP-TH.
Thus, since many available reviews on Lorentz violation are
masterly written\footnote{Among the many excellent reviews on the
ongoing search for Lorentz violation see, for instance, the one by
Mattingly \cite{mattingly} where an effort is made towards
including all theoretical approaches as well as the
phenomenological analysis.} we shall try here to do the reviewer's
job in a different perspective by trying to spell out some basic
issues behind the search for Lorentz violation.

\section{A bird's-eye-view on the search for Lorentz violation}

It is true that way after the experimental establishing of STR
there have been various attempts to modify aspects of STR in
various ways. Already Heisenberg questioned the causality
structure of Quantum Field Theory (QFT) while facing the problem
of the very definition of elementary particle \cite{heisenberg}.
On the other hand - although the paper that probably pioneered the
field now known as ``Lorentz violation'' is  that of 1990 by
Carroll, Field and Jackiw\footnote{There the authors introduced a
Lorentz violating term of Chern-Simons origin into Maxwell
electrodynamics and extensively studied the possibility of visible
signatures.} \cite{jackiw} - the contemporary systematic assault
to Lorentz symmetry can be ascribed to results of String Theory
(ST).

The first result is Kostelecky and Samuel's discovery that in (an
open bosonic, and possibly other) ST(s) Lorentz symmetry can be
spontaneously broken due to the presence in the string (field)
lagrangian of cubic terms that include one scalar (the tachion)
and two tensor fields (of any rank) \cite{kosteleckysamuel}. The
resulting low-energy limit theory is what Colladay and Kostelecky
later called the Standard Model Extension (SME)
\cite{colladaykostelecky}. The SME is certainly the most widely
used theoretical frame within which Lorentz violating
phenomenological scenarios are investigated today (for a review
see, e.g., \cite{SMEreview}).

The second result is of Seiberg and Witten who discovered that a
particular low-energy limit of ST leads to certain noncommutative
(in the sense of spatiotemporal coordinates, e.g. $[x_{\mu} ,
x_{\nu}] = i \theta_{\mu \nu}$) field theories (NCFTs)
\cite{seibergwitten}. They proved that these theories, although
Lorentz symmetry violating, still keep standard gauge invariance.
This was done by providing a map that gives the noncommutative
nonlocal gauge field (transforming under the noncommutative gauge
transformations) in terms of the standard commutative local gauge
field (transforming under the usual gauge transformations) and of
the noncommutative parameters $\theta_{\mu \nu}$. This map is
known as the Seiberg-Witten (SW) map\footnote{It is surely
possible to construct gauge theories on noncommutative spaces such
that the SW map holds without direct reference to ST, as shown in
\cite{wess} where the authors consider $[x , \cdot ~]$ as a
derivative and make this derivative gauge covariant. What we want
to stress here, though, is the profound impact on the HEP-TH
community and beyond of ideas originated in ST: would the
Seiberg-Witten map be discovered without Seiberg and Witten?}.
This is probably the second most used theoretical frame to study
Lorentz violation\footnote{The quantum phase of NCFTs is still an
open issue: it was shown that novel infrared divergencies appear
in such theories \cite{minwallaseiberg}, but whether these are
unavoidable or not is still to be clarified. Lately it was argued
that such divergencies should not be present if noncommutativity
is implemented via a deformed (twisted) coproduct
\cite{balachandran}.} (for a very recent overview see for instance
the Introduction of \cite{rivelles}).

The SME and the NCFTs, both ``string-inspired'' models, are not
the only two known approaches to Lorentz violation available
today. We single out these two because they are the most widely
used and because their explicit goal is to propose that Lorentz
violation could be a fact of nature. Another available approach
where Lorentz violation is posited as a fact of nature, is that of
Double Special Relativity (DSR) proposed by Amelino-Camelia in
\cite{camelia} (see also \cite{magueijo}), based on {\it two}
invariants: the speed of light and an energy scale. An important
independent work is that of Coleman and Glashow that do not take
that view \cite{colemanglashow}. Their idea is to {\it test} STR
by studying what sort of new bounds can be obtained at high
energies using the hypothesis that Lorentz symmetry is not exactly
satisfied as a working technical tool. Others approaches also
exist \cite{mattingly}.

Let us now enter a bit more into the details of the SME and NCFTs.
As said, the SME is based on a mechanism of spontaneous breaking
of Lorentz symmetry within a higher dimensional string field
theory. ``Over there'' lagrangians contain interactions that
include tensor fields of any rank, hence an infinite number of
them is possible. Lorentz symmetry exactly holds - where by
``exact'' here we mean that the lagrangian and the vacuum
configuration are both symmetric - and similarly for gauge
symmetry and renormalizability. This is one of the magic things of
ST: such interactions would spoil gauge symmetry and
renormalizability in a four dimensional field theory, but for
higher dimensional strings this is not a problem. One such stringy
interactions is a cubic term with a scalar and a squared tensor
field. For stability the vacuum expectation value (vev) of this
scalar field cannot be zero, thus giving rise to a nonzero mass
for the tensor fields that this way have nonzero vevs. As the
tensors transform non trivially under the Lorentz group this
mechanism gives rise to a spontaneous breaking of such symmetry.
As for any spontaneous breaking, it is the vacuum configuration
that becomes not symmetric, while the lagrangian remains so. By
the usual mechanism of the shifting/redefinition of the fields
(the rescaling of the vacuum to the ``true'' vacuum) the vevs of
these fields, say $C^{(k)}_{\mu ... \nu}$, appear in the
lagrangian.

In the SME the breaking is chosen so that such coefficients
combine with the fields of the SM (and gravity) and their
derivatives. The Lorentz (and possibly CPT) violating terms {\it
added} to the SM terms have then the form
\begin{equation}\label{SME1}
C^{(k)}_{\;\;\;\;\mu ... \nu} ({\rm SM \; fields \; and \;
derivatives})^{\mu ... \nu} \;,
\end{equation}
where $C^{(k)} \sim 1/m_{\rm Planck}^k$, so that the higher we go
with the power $k$, the smaller the effect. Notice here the
appearance of the fundamental length $\ell_{\rm Planck} \sim
m^{-1}_{\rm Planck}$.

The idea here is to find phenomenological setups such that the
``Planck size'' effects can be ``magnified'' to the point of
giving rise to signatures measurable by the instruments at our
disposal. Those effects should have visible signatures ``over
here'', i.e. at reachable energy scales too. This is quite an
interesting point of view. It is somehow opposed to the idea that
it is impossible to test phenomena that take place at the Planck
scale. Of course, it is not claimed that this way one reaches the
``Planck world'', but the core of the approach, so to speak, is
that the ``Planck world'' and ``our world'' cannot be fully
decoupled.

It is crucial to notice that the Lorentz indices match: the
Lorentz violating terms do not violate the symmetry explicitly
(as, for instance, a non-matching index would brutally do) but
they are still scalars when {\it all} the terms, the $C^{(k)}$s
and the SM fields and derivatives, are transformed. On the
contrary, when the $C^{(k)}$s are {\it not} transformed (as one
would expect, for instance, if these terms would represent a fixed
background) then Lorentz violation comes about. This occurrence is
usually referred to as: The SME is ``observer-Lorentz'' invariant,
while it is ``particle-Lorentz'' violating, the ``observer''
transformations being those of the first kind - i.e. those
involving the coefficients {\it and} the fields - the ``particle''
transformations being those of the second kind.

This model brings with it many nice features of the original ST
model: it is powercounting renormalizable, gauge invariant,
$P_\mu$ is conserved and $P_0 > 0$. These properties made SME's
success and the various $C^{(k)}$s for the different sectors of
the SME (especially those in the electromagnetic sector) have been
cleverly searched for in a great variety of experiments and
experimental proposals (see e.g. \cite{SMEreview}).

We want to discuss now the example of the spectrum of the
synchrotron radiation to give a flavor of the kind of
modifications one encounters by dealing with Lorentz violation and
to make explicit one of the mechanisms by which ``amplification''
of Planck-size effects can take place. We shall not present the
SME treatment but instead we shall treat the case within a
noncommutative electrodynamics (NCED) approach\footnote{To a
certain extent NCFTs can be seen as a SME with a particular choice
of the coefficients $C^{(k)}_{\mu ... \nu}$ \cite{mattingly},
\cite{SMEreview},\cite{smeandncft}.}. This will also give us the
chance to introduce some details on the basic formalism of NCFTs
approach.

Suppose that, due to a fundamental length $\ell_{\rm string} \sim
\sqrt{\theta} \sim \ell_{\rm Planck}$, spatiotemporal coordinates
do not commute. One way of making this explicit is\footnote{Of
course, this is not the most general way noncommutativity of the
coordinates could take place. For instance, two equally valid, if
not more general, approaches are the Lie-algebraic and the
coordinate-dependent ($q$-deformed) formulations \cite{wess}, and
many other approaches exist. Nonetheless, the canonical form is
surely the most simple and the basic features of noncommutativity
are captured in this model.}
\begin{equation}  \label{1}
x^\mu * x^\nu - x^\nu * x^\mu = i \theta^{\mu \nu} \;,
\end{equation}
where the (Moyal-Weyl $*$-)product of any two fields $\phi(x)$ and
$\chi(x)$ is defined as $(\phi * \chi) (x) \equiv e^{ \frac{i}{2}
\theta^{\mu \nu} \partial^x_\mu \partial^y_\nu} \phi (x) \chi
(y)|_{y \to x}$, and it is a suitable generalization of the
multiplication law in the presence of a nonzero $\theta$. Finally,
$\theta^{\mu \nu}$ is $c$-number valued, constant, and the Greek
indices run from $0$ to $3$. The NCED action is
\begin{equation}  \label{ncym} \hat{I} = -
\frac{1}{4} \int d^4 x \hat{F}^{\mu \nu} \hat{F}_{\mu \nu} \;,
\end{equation}
where $\hat{F}_{\mu \nu} = \partial_\mu \hat{A}_\nu - \partial_\nu
\hat{A}_\mu - i [ \hat{A}_\mu , \hat{A}_\nu ]_* $. As said, the
nonlocal field $\hat{A}_\mu$ can be expressed in terms of a
standard U(1) gauge field $A_\mu$ and of $\theta^{\mu \nu}$ by
means of the SW map\footnote{$\hat{A}_\mu (A,\theta) \to A_\mu$ as
$\theta^{\mu \nu} \to 0$, hence, in that limit, $\hat{F}_{\mu \nu}
\to F_{\mu \nu} = \partial_\mu A_\nu - \partial_\nu A_\mu$.}
\cite{seibergwitten} $\hat{A}_\mu (A,\theta)$, that at $O(\theta)$
reads
\begin{equation}
\hat{A}_{\mu}(A, \theta) = A_{\mu} - \frac{1}{2} \theta^{\alpha
\beta}A_{\alpha}(\partial_{\beta}A_{\mu} + F_{\beta \mu})  \;.
\end{equation}

The Noether currents for space-time transformations of the theory
(\ref{ncym}) are $J_f^\mu = \Pi^{\mu \nu} \delta_f A_\nu -
\hat{\cal L} f^\mu$, where $\Pi^{\mu \nu} = \delta \hat{\cal L} /
\delta
\partial_\mu A_\nu$, and for translations (the only symmetric
case) $f^\mu \equiv a^\mu \equiv {\rm const}$. Hence the
(conserved) energy-momentum tensor $T^{\mu \nu} = \Pi^{\mu \rho}
F^\nu_\rho - \eta^{\mu \nu} \hat{\cal L}$ is in general not
symmetric, a clear sign of the breaking of Lorentz symmetry
\cite{iorio}. Conservation of $T^{\mu \nu}$ means conservation of
the Poynting vector
\begin{equation}\label{poynting}
  \vec{S} = \frac{c}{4 \pi} \vec{D} \times \vec{B} =
  \frac{c}{4 \pi} \vec{E} \times \vec{H} \;,
\end{equation}
where $D^i \equiv \Pi^{i 0}$ and $H^i \equiv \frac{1}{2} \eps^{i j
k} \Pi_{j k}$ are the constitutive relations containing all the
relevant information about the Lorentz violating vacuum.

The effects of a nonzero $\theta$, if any, are very small, thus
the $O(\theta)$ model would do the job
\begin{equation}  \label{othetamaxwell}
\hat{I} = - \frac{1}{4} \int d^4 x \; [F^{\mu \nu} F_{\mu \nu}
-\frac{1}{2} \theta^{\alpha \beta} F_{\alpha \beta} F^{\mu \nu}
F_{\mu \nu} + 2 \theta^{\alpha \beta} F_{\alpha \mu} F_{\beta \nu}
F^{\mu \nu}] + J_\mu \hat{A}^\mu \;,
\end{equation}
where the vector field is coupled to an external current, and the
$O(\theta)$ SW map and $*$-product are used. In the presence of a
background magnetic field $\vec{b}$, and for $J_\mu = 0$, the
plane-wave solutions exist \cite{jackiw2}. The waves propagating
transversely to $\vec{b}$ travel at the modified speed $c'=c(1 -
\vec{\theta}_T \cdot \vec{b}_T)$ (where $\vec{\theta} \equiv
(\theta^{1}, \theta^{2}, \theta^{3})$, with $\theta^{i j} =
\eps^{i j k} \theta^{k}$, and $\theta^{0 i} = 0$) while the ones
propagating along the direction of $\vec{b}$ still travel at the
usual speed of light $c$.

Plane-waves do not represent a case where ``amplification'' of the
nonzero $\theta$ effects can take place. Synchrotron radiation,
instead, is a case where ultrarelativistic ``magnification'' of
the electromagnetic field in the direction of motion could help.
With these ideas in mind the authors of \cite{ciz} studied the
synchrotron radiation in NCED, and indeed found the amplification
they were looking for. It is matter of solving the modified
Maxwell equations descending from (\ref{othetamaxwell}) with the
settings: a) charged particle moving (circularly) in the plane
$(1,2)$: $J_\mu = e c \beta_\mu \delta(x_3) \delta^{(2)} (\vec{x}
- \vec{r}(t))$, where $\vec{r}(t)$ is the position of the
particle, and $\beta_\mu = (1, \vec{v}/c)$; b) $\vec{b} = (0, 0,
b)$; c) $\vec{\theta} = (0, 0, \theta)$. Keeping only
contributions of order $O(e/R)$, where $R$ is the distance from
the source, one can compute the electric and magnetic fields that
have quite involved expressions\footnote{The part proportional to
$\lambda \equiv 2 \theta b$ contains a term of the form
\[
  \left[\frac{1}{c (1 - \vec{n} \cdot
\vec{\beta})} \frac{d}{d t'} \left( \frac{1}{c (1 - \vec{n} \cdot
\vec{\beta})} \frac{d}{d t'} \frac{\vec{n} c (t - t')}{(1 -
\vec{n} \cdot \vec{\beta}) R} \right) \right]_{\rm ret} \;,
\]
where $\vec{n} = \vec{R} / R$, and $[\quad]_{\rm ret}$ are the
usual retarded quantities. We see here contributions proportional
to the {\it derivative} of the acceleration. The $\ddot \beta$
contribution arises as an effect of the conversion of the two
speeds  of lights $c$ and $c'$ in the poles of the Green functions
into a single speed $c$ with a {\it derivative} of the delta
function $\delta'(\tau - R /c)$. These terms recall the familiar
scenario of the Abraham-Lorentz pre-acceleration effects for the
classical self-energy of a point charge \cite{jackson}.}.

To compute the spectrum in the ultra-relativistic regime ($\beta =
v / c \to 1$) and far from the source ($|\vec{x}| \sim R >>
  |\vec{r}(t)|$) one needs the power radiated in the direction $\vec
  n$: $ d P (t) / d \Omega = R^2 [\vec{S} \cdot \vec{n}] \equiv |\vec{L} (t)|^2$,
where all the quantities ($\vec{n}, \vec{\beta}$,
$\dot{\vec{\beta}}$, $R$) are in the plane $(1,2)$, and the
Poynting vector $\vec{S}$ given in (\ref{poynting}) contains all
the relevant information. In the ultra-relativistic approximations
two are the characteristic frequencies for the synchrotron: the
cyclotron frequency $\omega_0 \sim c / |\vec{r}|$, and the
critical frequency $\omega_c = 3 \omega_0 \gamma^3$. For radiation
in the plane the relevant range of frequencies is $\omega >>
\omega_0$ (such that the latitude $\vartheta \sim \pi /2$), hence
the leading terms for the energy $d I (\omega) / d \Omega \equiv 2
|\vec{L} (\omega)|^2$ (where $\vec{L} (\omega)$ is the Fourier
transformed of $\vec{L} (t)$) are
\begin{equation}\label{energy2} \frac{d}{d \Omega} I
(\omega) \sim \frac{e^2}{3 \pi^2 c} \left( \frac{\omega}{\omega_0}
\right)^2 \gamma^{-4} \left[ K_{2/3}^2 (\xi) [1 + \lambda (1 + 6
\gamma^2)] + \lambda \frac{24 \gamma^5 \omega_0}{\omega} K_{1/3}
(\xi) K_{2/3} (\xi) \right] \;,
\end{equation}
where $\xi = (\omega / 3 \omega_0) \gamma^{-3}$, and $K_{2/3}
(\xi)$ and $K_{1/3} (\xi)$ are modified Bessel
functions\footnote{When $\omega << \omega_c$, for $\xi \to 0$,
$K_{\nu} (\xi) \sim \xi^{- \nu}$, $\nu = 2/3 , 1/3$.}. For
$\omega_0 << \omega << \omega_c$, i.e. $1 << \omega / \omega_0 <<
\gamma^3$,
\begin{equation}\label{ratio}
X \equiv \frac{d I (\omega) / d \Omega}{d I(\omega) / d
\Omega|_{\lambda = 0}} \sim 1 + 10 (\omega_0 / \omega)^{2/3}
\lambda \gamma^4 \;.
\end{equation}
This shows how the smallness of the factor $\lambda = 2 \theta b$
is compensated by the fourth power of $\gamma$, hence the
amplification one was looking for. For instance, by using the
bound in \cite{bound}, $\theta < 10^{-2} ({\rm TeV})^{-2}$, one
has that $\lambda = 2 b \theta < 2 n 10^{-23}$, where $n$ is the
value of $b$ in Tesla, and 1 Tesla $\sim 10^{-21} ({\rm
TeV})^{2}$. Thus for an electron synchrotron the correction is
\begin{equation}
X < 1 + (\frac{\omega_0}{\omega})^{2/3} n \times 10^{-21} \times
\left( \frac{{\cal E} ({\rm MeV})}{{\rm MeV }}\right)^{4} \;,
\end{equation}
where $\cal E$ is the energy of the electron, $\gamma_{\rm max}
\sim 2 {\cal E} ({\rm MeV}) / {\rm MeV}$. For the most energetic
synchrotron (SPring-8, Japan) ${\cal E} = 8$ GeV, $b \sim 1$
Tesla, and when $\omega / \omega_0 \sim \gamma^2$ we
have\footnote{These numbers should be taken with some care, as the
scope here is to illustrate the mechanism of amplification of
Lorentz violating effects at work rather than to give the most
severe bounds. We are also justified in doing so here by the fact
that within NCED, at present, no better constraints are obtained
by including synchrotron radiation of astrophysical origin (and
vacuum \v{C}erenkov radiation) \cite{ciz2}.}
\[
X  < 1 + 10^{-10} \,.
\]
We see a ``$10^{13}$-amplification'' of the effects induced by a
nonzero $\theta$, independent from the actual input value for
$\theta$: from $10^{-23}$ to $10^{-10}$ in this case.

Other investigators have also looked into the synchrotron
radiation within the SME \cite{altschul} and other approaches
\cite{urrutia}. They find similarly important departures from the
Lorentz preserving formula in agreement with the above presented
modification of the energy spectrum and with a previously proposed
formula for the maximum frequency based on kinematical general
arguments \cite{jacklibmat}.

Synchrotron radiation is just one among the many phenomenological
setups (a large part of which in the electromagnetic sector) that
have been lately investigated, and we make here no claim that it
is the most important. We picked up this example because it
clarifies the amplification mechanism quite cleanly and because of
our own familiarity with it. Despite the growing number of
phenomenological/experimental investigations, no signals of
violation are found. In the most fortunate cases, the best we are
able to do is to use existing data to ameliorate the bounds on the
violating parameters. The last work of Bailey and Kostelecky
\cite{baileykostelecky} is dedicated to the pure gravity sector of
the SME and many experimental setups to look for Lorentz violation
are proposed. In a way, that paper can be seen as a turning point.
The message there seems to be ``As the electromagnetic sector is
giving no positive results, let us now turn our attention to the
gravity sector''.

To our knowledge, this is the state of the art. The reviewer's job
could safely stop here but, while doing it, we had the chance to
take a step back from the everyday work-schedule and we could take
a look at the {\it grand scenario} - it is of Lorentz symmetry
that we are talking here - thus some general considerations
naturally came to the mind. The rest of this lecture is dedicated
to that.

\section{Three Questions}

Here are three questions:

\subsection{Why questioning a solid paradigm?}

Why are we so systematically ``shaking'' the STR? There are only
two kinds of reasons to embark oneself to turn inside out a
solidly tested view of nature: theoretical or experimental. The
first case (theory) reduces to the second because, although the
motivations to change can be fully theoretical, in the end the new
theory has to face the experiments.

When we learned the STR we also learned how compelling were the
{\it experimental} evidences that forced theorists to re-think the
Galilean/Newtonian paradigm. Today this is not the
case\footnote{Another way of looking at the historical development
of STR is to say that a {\it wrong} theoretical approach (the
Galilean/Newtonian) pushed for experimental findings, such as that
of a luminiferous aether, that {\it never came}. These {\it
non-findings} where the experimental basis on which STR was
founded. When put it this way the sentence in the text above
should end with a question mark.}. On the other hand, looking
closely, the argument of having compelling {\it theoretical}
motivations presents a peculiarity that deserves consideration.
The logic is as follows: If a theory claims to describe nature
than it must come under the judgment of the experiments. It must
provide predictions for precise signatures so that the theory
itself can be proved or disproved. The SME, the NCFTs and the
other approaches to Lorentz violation are surely taking the road
of experiments, there is no doubt about that and it is a most
welcome fact. But, in our view, the underlying motivations behind
the search for Lorentz violation are in ST. This is more evident
for the SME or for the NCFTs {\it \`{a} la} Seiberg-Witten, but
more loosely speaking it can be said of the other approaches too.
This leads straight to our answer to the question ``why
questioning STR'': we are questioning it because the dominant
theory at our disposal (ST) is {\it somehow} telling us to do
that. The reason why the (deep) theoretical motivations behind the
search for Lorentz violation have a logical loophole is that: if
{\it no signs} of violations are found this would {\it not} affect
ST. In the negative case the SME would probably be left apart, but
nothing at all would happen to ST.

It is true that, strictly speaking, ST is not necessary to write
down the SME, as explicitly shown by its inventors Colladay and
Kostelecky \cite{colladaykostelecky}, and by many authors
\cite{SMEreview}. But, while this has a pleasant outcome as for
the model independence and generality of certain results, on the
other hand it is clear that the motivations to build up the model
in the first place lay with ST. Similarly, the whole path from
noncommuting coordinates till noncommutative gauge theories can be
taken without mentioning the word ``string'' (see, e.g.,
\cite{wess}). Nonetheless, it was the link to ST provided by
Seiberg and Witten \cite{seibergwitten} that made NCFTs' fortune
as viable candidates for Lorentz violation. In a more loose sense
it can be said that also the other approaches are permeated by
concepts and ideas originated in or promoted by ST (such as a
fundamental length/energy scale and nonlocality).

Thus, our answer to the question immediately turns on a lightspot
on the very special status of ST: it leads the research in
particle physics by inspiring, one way or the other, the new
directions of investigation that, sometimes, reach the arena of
experimentally testable scenarios, but ST itself never gets so
involved with its ``by-products'' to the extent of being proved
false\footnote{It is expected that, for consistency, positive
outcomes of experiments, for instance in Lorentz violation, will
not be taken as a proof (direct or indirect) for the experimental
soundness of ST.}. This is very singular, and we do not know of
any other case like that in science. To use a simplifying image:
it is as if ST gets {\it close} to the border between theory and
phenomenology, leaves there the results of its speculations and
leaves to others to write down things in a form suitable for
testing, but then these models are {\it no longer part} of ST.
Another question then comes out of the first question: is ST a
{\it physical} theory? This is a very delicate (or rough?)
question to ask. We should soon try some speculations. Before
doing so let us close this subsection by further clarifying our
own view on the search for Lorentz violation.

Despite the lack of experiments and despite the absence of
compelling theoretical reasons, for us this search still
represents an important chance for a fresh rethinking of
fundamental issues in particle physics - the most important of all
being the symmetry principles. This way, if we are lucky, we might
discover by unsuspected means where and if the ``mistake'' is
hiding. By ``mistake'' here we mean that prejudice we are
enforcing on our view that lead us too ``far away''. It is a
particularly welcome circumstance that is the community of
theorists and experimentalists together that is facing these
issues.

\subsection{Is ST a physical theory?}

This is a difficult question. We could measure how difficult it is
by showing how spread between ``Yes'' and ``No'' is the plot of
the answers physicists would give. There would be straight ``Of
course it is not'', and straight ``Of course it is''. Most
interestingly, there would be a blur phase of answers where the
word ``physics'' would start changing its meaning from the
original one - i.e. of an experimental science as we learned it at
school - towards something less and less definite. This last class
of answers (containing the ``Yes'') could be the starting point
for a serious discussion involving scientists and philosophers
that, in our opinion, is much needed today. One way of attacking
the problem (because this is a problem) would be to invoke
Bridgman's operationism \cite{bridgman} and see in which of its
class ST would fit. Needless to say, according to Bridgman's
criteria, ST is not a physical theory. Nonetheless, more care
would be needed because Bridgman's criteria are surely not
infallible and they were invented in a different era of particle
physics. On top of that, we are only amateur philosophers hence
unable to discern among the subtleties of {\it logical coherence}
sometimes invoked by ST to justify its existence. Thus, we are
getting too far away without a definite answer. Let us try
something else.

Let us try attacking the problem by talking about a {\it string
theorist}'s\footnote{I was inspired in doing so by David Tong (at
that time a fellow postdoc at MIT) who used to say something like
``My goal is to eventually obtain an experimental prediction from
string theory or, at the very latest, from a {\it string
theorist}.''} trajectory: Alan Kostelecky's. To many Kostelecky is
the man of Lorentz violation. He started off his career as a
string theorist but of a special kind. Reading the papers he
co-wrote on Lorentz violation in ST and the seminal ones on the
SME one clearly notices that even before moving fully into the
realm of experimentally verifiable (or falsiable) ideas he had the
(sane) attitude of a theorist with an eye (or even two) to the
experimental arena.

There is a - probably unnoticed - sentence in the seminal paper on
the SME that brings us a step forward towards the answer we are
looking for: In the Introduction of the second entry of Reference
\cite{colladaykostelecky} it is described in words the mechanism
of spontaneous breaking of Lorentz symmetry within the higher
dimensional ST. There Kostelecky and Colladay call the four
space-time dimensions {\it physical}. This implies that the other
(twentytwo or seven or whatever) dimensions are considered there
{\it un-physical}. So here we have at least part of the answer to
our question (``Is ST a physical theory"): according to a string
theorist that part of string theory that studies ``phenomena''
outside the four dimensions is not physics\footnote{As a
byproduct, we also solved Tong's problem of obtaining experimental
predictions from a string theorist: ask Alan Kostelecky.}.

To make full justice to Bridgman we should ask the same question
we asked for ST for other theories/concepts available today in
theoretical particle physics. These include Suypersymmetry (SUSY)
and its descendants, Extra Dimensions, Noncommutativity {\it \`{a}
la} Connes, Loop Quantum Gravity, not to talk about models
nowadays out of fashion such as the Grand Unification models, etc.
Surely, the discussion among scientists and philosophers we evoked
earlier should scrutinize those ideas as well. Nonetheless, we
make three points here: First, ST is the {\it dominant} theory (as
we think we proved in the case in point of the search for Lorentz
violation) and as any good regent has to fully take the burden
that goes with the honors. Second, most of (or all) the concepts
and ideas above enumerated are somehow included into the ``all
encompassing'' ST. Third, we were motivated by our investigations
of the philosophical/epistemological reasons behind the search for
Lorentz violation and we simply found ST behind it.

\subsection{How about SUSY?}

SUSY surely has its place among the ideas/theories waiting for an
answer about its physical relevance. Nonetheless, physicists have
to move fast and before the problem is fully solved (it might take
forever...) one might try to take shortcuts. As said before, in
our view, the search for Lorentz violation is very welcome exactly
because it might serve the scope of changing in an unsuspected
fashion ideas that were on the table. One of these ideas could be
SUSY.

Already Berger and Kostelecky \cite{susykostelecky} have
investigated, within the SME, the effects of Lorentz violation on
SUSY and found that a (modified) SUSY algebra can be written.
Similarly, the effects of noncommuting coordinates on SUSY has
been considered \cite{susyrivelles}. What if we take a step back
and look into the no-go theorems \cite{colemanmandula} with the
assumption that the spatiotemporal group is no longer ISO(3,1)?

\vspace{.5cm}

\noindent {\large \bf Acknowledgments}

It is my pleasure to thank the organizers of DICE2006, especially
Thomas Elze, for inviting me. I also thank Ugo Senatore of the
National Monument Library of Badia di Cava de' Tirreni in Italy,
for providing an inspiring environment for the preparation of this
lecture.

\end{document}